\documentclass[12pt,a4paper,dvips]{article}

\usepackage{graphicx}
\usepackage{amsbsy}
\usepackage{amssymb}

\begin{document}

\newcommand{\be}{\begin{equation}}
\newcommand{\ee}{\end{equation}}
\newcommand{\bea}{\begin{eqnarray}}
\newcommand{\eea}{\end{eqnarray}}
\newcommand{\tr}{\,\hbox{tr }}
\newcommand{\Tr}{\,\hbox{Tr }}
\newcommand{\Det}{\,\hbox{Det }}
\newcommand{\fslash}{\hskip-.2cm /}
\begin{center}
{\bf \Large Improved Gaussian Approximation}
\end{center}
\vspace*{0.2cm}
\begin{center}
A. Bala\v z, A. Beli\'c and A. Bogojevi\'c
\end{center}
\begin{center}
{\it Institute of Physics\\
P.O.B. 57, Belgrade 11001, Yugoslavia}
\end{center}
\vspace*{0.1cm}
\begin{abstract}
{\small In a recently developed approximation technique \cite{ab3}
for quantum field theory the standard one-loop result is used as a
seed for a recursive formula that gives a sequence of improved
Gaussian approximations for the generating functional.
In this paper we work with the generic
$\phi^3+\phi^4$ model in $d=0$ dimensions. We compare the first,
and simplest, approximation in the above sequence with the one-loop and
two-loop approximations, as well as the exact results
(calculated numericaly).}
\end{abstract}

The central object in quantum field theory is the generating
functional $Z[J]$. Functional derivatives of $Z[J]$ with respect to the
external fields $J(x)$ give the Green's functions of the theory. The
generating functional is determined from the (Euclidian) action $S[\phi]$
through the path integral
\be
Z[J]=\int [d\phi]\,e^{-\left(S[\phi]-\int dx\,J(x)\,\phi(x)\right)}\ .
\ee
The integration measure is, formaly, simply
$[d\phi]=\prod_{x\in\mathbb{R}^d}d\phi(x)$,
where $d$ is the dimension of space-time. In this paper we will work with
models in $d=0$ dimensions. In $d=0$ functionals become functions,
and the path integral reverts to a single definite integral over the whole
real line
\be
Z(J)=\int d\phi\,e^{-\left(S(\phi)-J\,\phi\right)}\ .\label{z}
\ee
Two further important objects are $W(J)$ --- the generator of connected
diagrams (or free energy)
\be
Z(J)=Z(0)\,e^{-W(J)}\ ,
\ee
and the quantum average of the field
$\varphi=\langle\phi\rangle=-\,\frac{\partial}{\partial J}\,W(J)$.

In the Gaussian approximation, we Taylor expand the action in the path
integral around some reference point $\phi_\mathrm{ref}$, and keep terms
that are at most quadratic in $\phi-\phi_\mathrm{ref}$. The integral in
(\ref{z}) is now a Gaussian and we find
\be
W_\mathrm{Gauss}(J,\phi_\mathrm{ref})=S(\phi_\mathrm{ref})-
J\,\phi_\mathrm{ref}+
\frac{1}{2}\,\ln S''(\phi_\mathrm{ref})-
\frac{1}{2}\,\frac{(S'(\phi_\mathrm{ref})-J)^2}{S''(\phi_\mathrm{ref})}\ .
\label{w-g}
\ee
For this approximation to make sense, the integral must get its dominant
contribution from the vicinity of the reference point $\phi_\mathrm{ref}$.
The standard Gaussian approximation (loop expansion) corresponds to the
choice $\phi_\mathrm{ref}=\phi_\mathrm{class}\,(J)$, where
$\phi_\mathrm{class}$ is the solution of the classical equation of motion

$S'=J$. The classical solution is the maximum of the integrand in
(\ref{z}). 

\begin{figure}[!ht]  
    \centering
    \includegraphics[height=5.5cm]{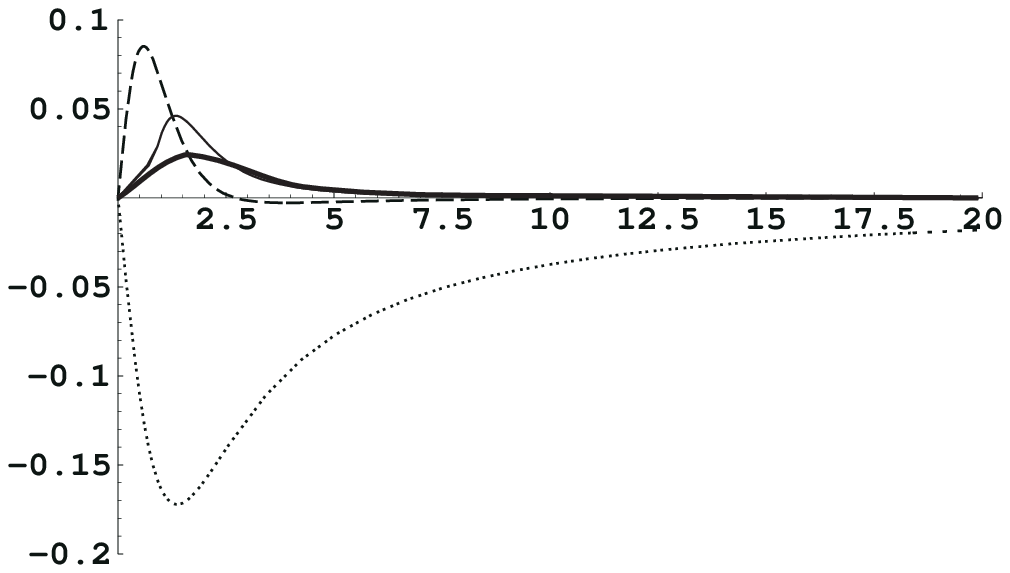}
    \caption{Plots of $\varphi-\varphi_0$ (dotted line),
    $\varphi-\varphi_1$ (dashed line), $\varphi-\varphi_2$ (thin line)
    and $\varphi-\varphi_\infty$ (thick line) as functions of $J$.
    The action is given in (\ref{action}) with couplings $g_3=0$, $g_4=1$.}
    \label{4k}
\end{figure}

In a previous paper \cite{ab3} we expanded the integrand around the average
field $\varphi$. As we have shown, although the classical solution gives
the maximum of the integrand, expansion around $\varphi$ gives a better
approximation for the area under the curve. The Gaussian approximation
around the average field $\varphi$ is simply
\be
W_\mathrm{Gauss}(J,\varphi)=S(\varphi)-J\,\varphi+
\frac{1}{2}\,\ln S''(\varphi)-
\frac{1}{2}\,\frac{(S'(\varphi)-J)^2}{S''(\varphi)}
\ .\label{w_q}
\ee
To be able to calculate this \emph{in closed form} we need to know
$\varphi(J)$, which is tantamount to knowing how to do the theory exactly,
since $\varphi$ and its derivatives give all the connected Green's functions.
The use of equation (\ref{w_q}) comes about when one solves it iteratively.
Using the definition of $\varphi$ in terms of $W$, as well as equation
(\ref{w-g}) we obtain the following iterative process
\be
\varphi_{n+1}(J)=-\,\frac{d}{dJ}\,W_\mathrm{Gauss}(J,\varphi_n(J))\ .
\label{phi-n}
\ee
For the seed of this iteration we chose the classical field, i.e.
$\varphi_0=\phi_\mathrm{class}$. In this way one obtains a sequence of
points $\varphi_0,\varphi_1,\varphi_2,\ldots$ or equivallently of
approximations to the connected generating functional
$W_1,W_2,W_3,\ldots$ given by $W_{n+1}(J)=W_\mathrm{Gauss}(J,\varphi_n(J))$.
In \cite{ab3} we have shown that this sequence gives better and better
approximations and converges (though slowly) to the best Gaussian
approximation $\varphi_\infty$. This is shown in Figure~\ref{4k}.
Note that $\varphi_\infty$ is still not  equal to the exact result $\varphi$.
The reason for this is obvious: We used the Gaussian approximation
$W_\mathrm{Gauss}$ in defining our recursive relation, and there is no reason
to expect that this converges to the exact result. 

So far we have seen that we can improve on the usual loop
expansion. However, the sequence of improved Gaussian approximations
converges very slowly. For this reason it is interesting to look at the
first approximations in this sequence and compare them to standard
approximation schemes. As we have seen, $\varphi_0=\phi_\mathrm{class}$,
so $W_1(J)$ is just the one-loop result. Our first new approximation is
therefore $W_2(J)=W_\mathrm{Gauss}(J,\varphi_1)$, where
$\varphi_1=-\frac{d}{dJ}W_1$. The nice thing about this approximation is
that it is only a bit more complicated than the one-loop result. From
now on we will study this approximation, and compare it to one-loop and
two-loop results, as well as to the exact results that have been calculated
numericaly. In what follows we will designate $W_2$ as the improved Gaussian
approximation.

For our comparison we have looked at the (full) $n$-particle Green's
functions
\be
G_n=\frac{\int d\phi\,\phi^n\,e^{-S(\phi)}}
{\int d\phi\,e^{-S(\phi)}}\ .
\ee
The model we considered was
\be
S(\phi)=\frac{1}{2}\,\phi^2+\frac{1}{3!}\,g_3\,\phi^3+
\frac{1}{4!}\,g_4\,\phi^4\ .\label{action}
\ee

\begin{figure}[!ht]  
    \centering
    \includegraphics[height=7cm]{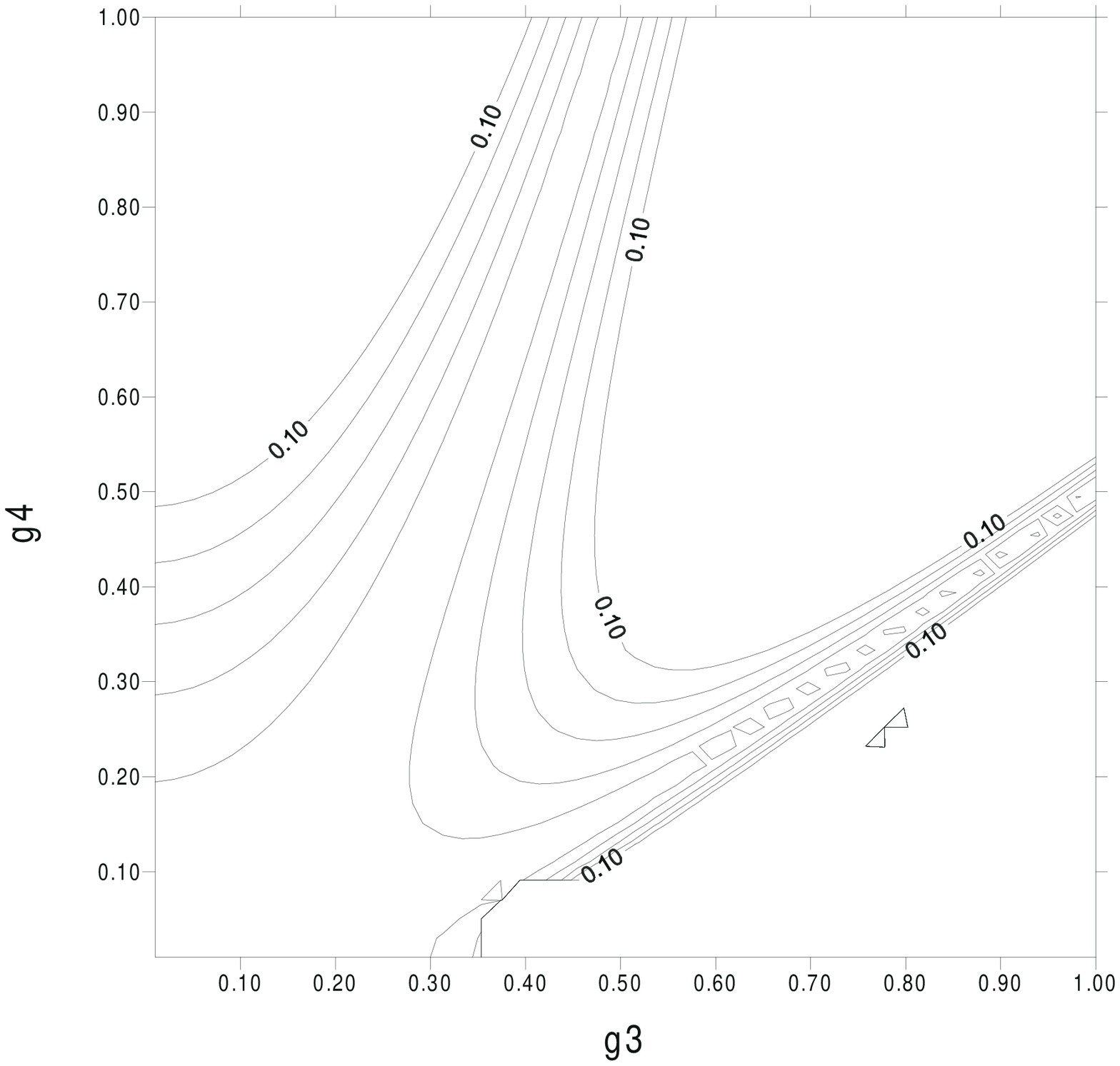}
    \caption{Absolute values of relative errors of
    the one-loop approximation to $G_2$. Only the contours corresponding
    to $|\delta G_2^\mathrm{1-loop}|<0.1$ are shown.}
    \label{l1g2}
\end{figure}

\begin{figure}[!ht]
    \centering
    \includegraphics[height=7cm]{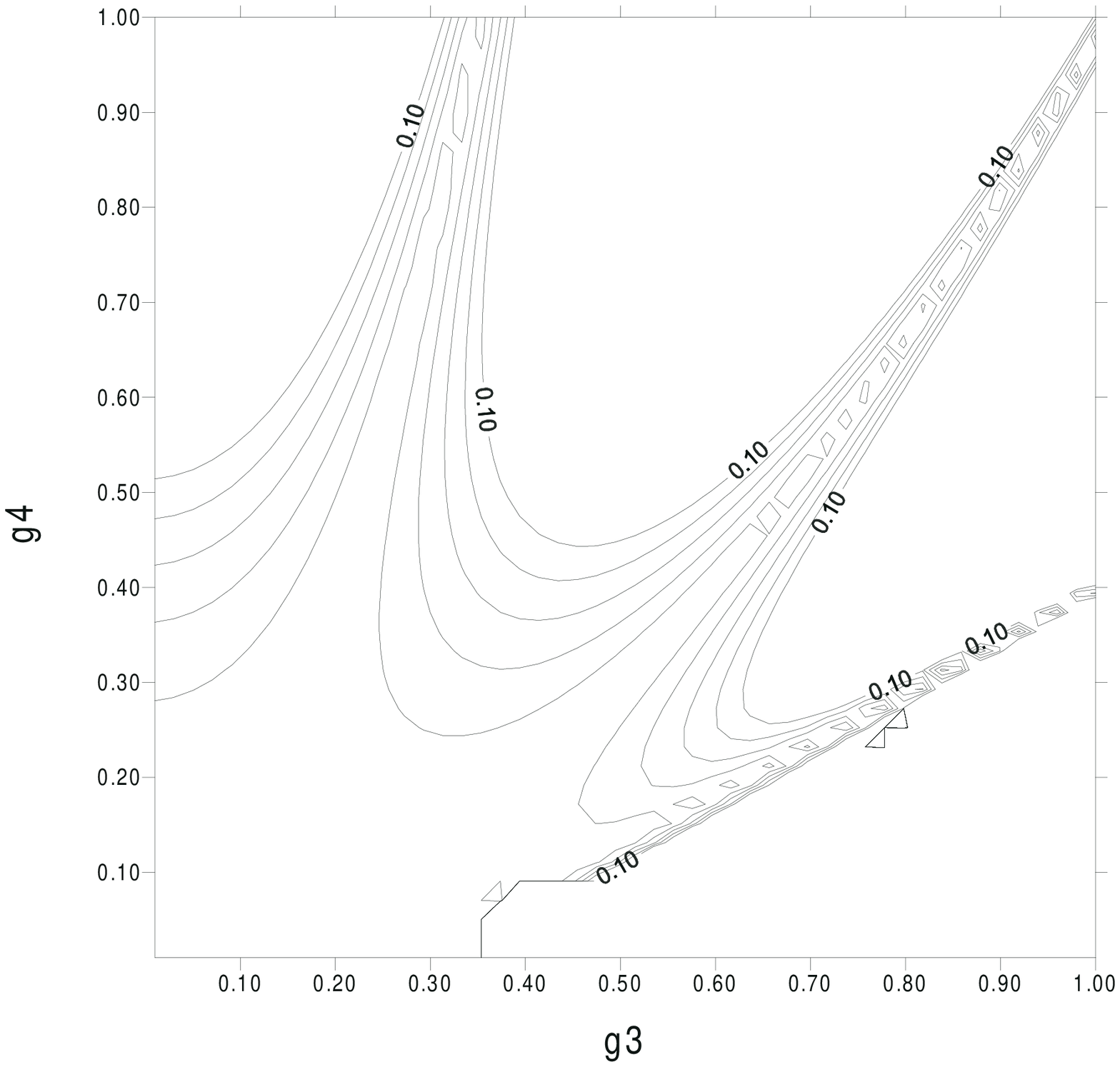}
    \includegraphics[height=7cm]{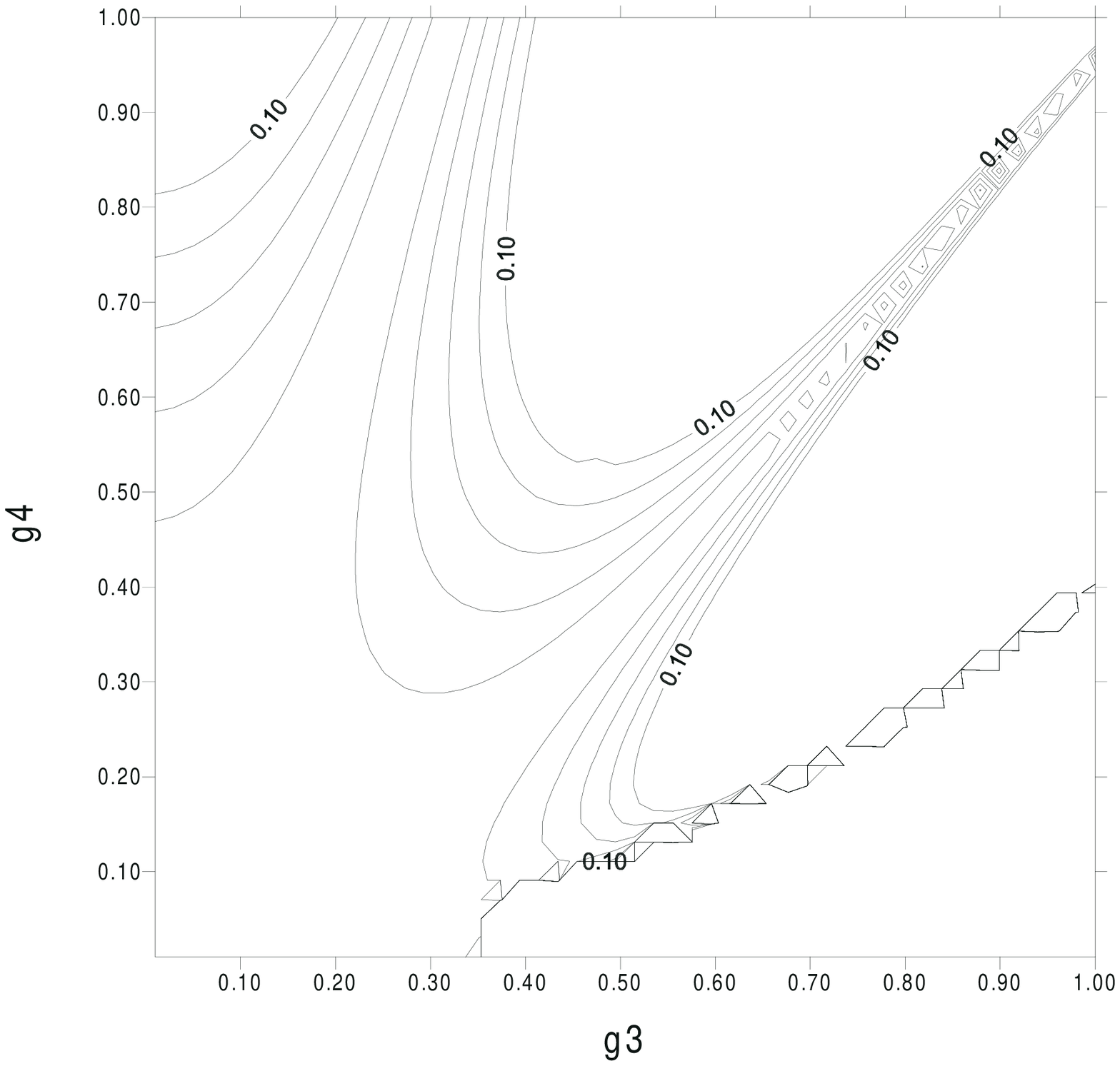}
    \caption{Contour plot of $|\delta G_2^\mathrm{2-loop}|<0.1$ (left), and
    $|\delta G_2^\mathrm{improved}|<0.1$ (right)}
    \label{l2g2+ig2}
\end{figure}

\begin{figure}[!ht]
    \centering
    \includegraphics[height=7cm]{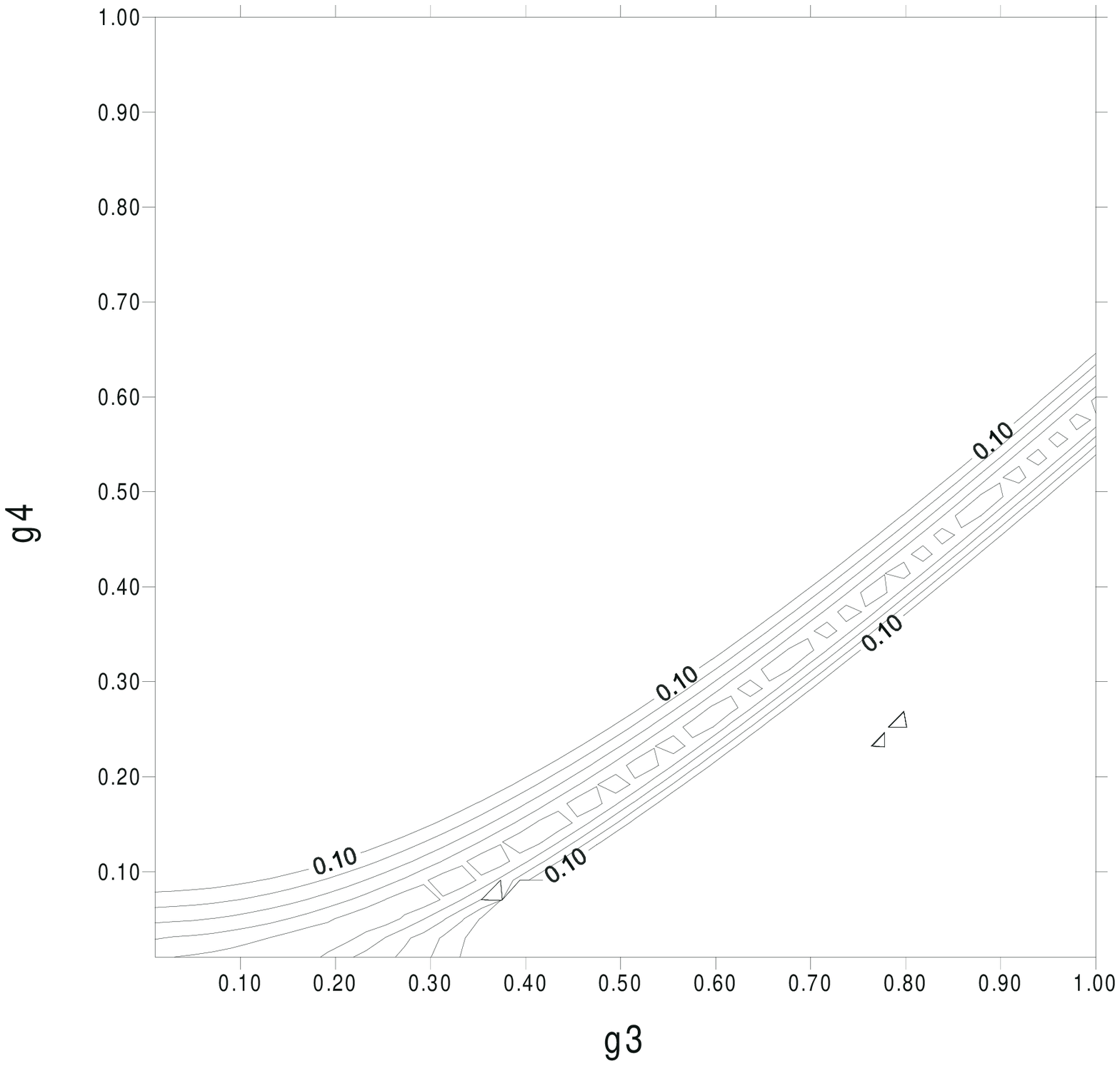}
    \includegraphics[height=7cm]{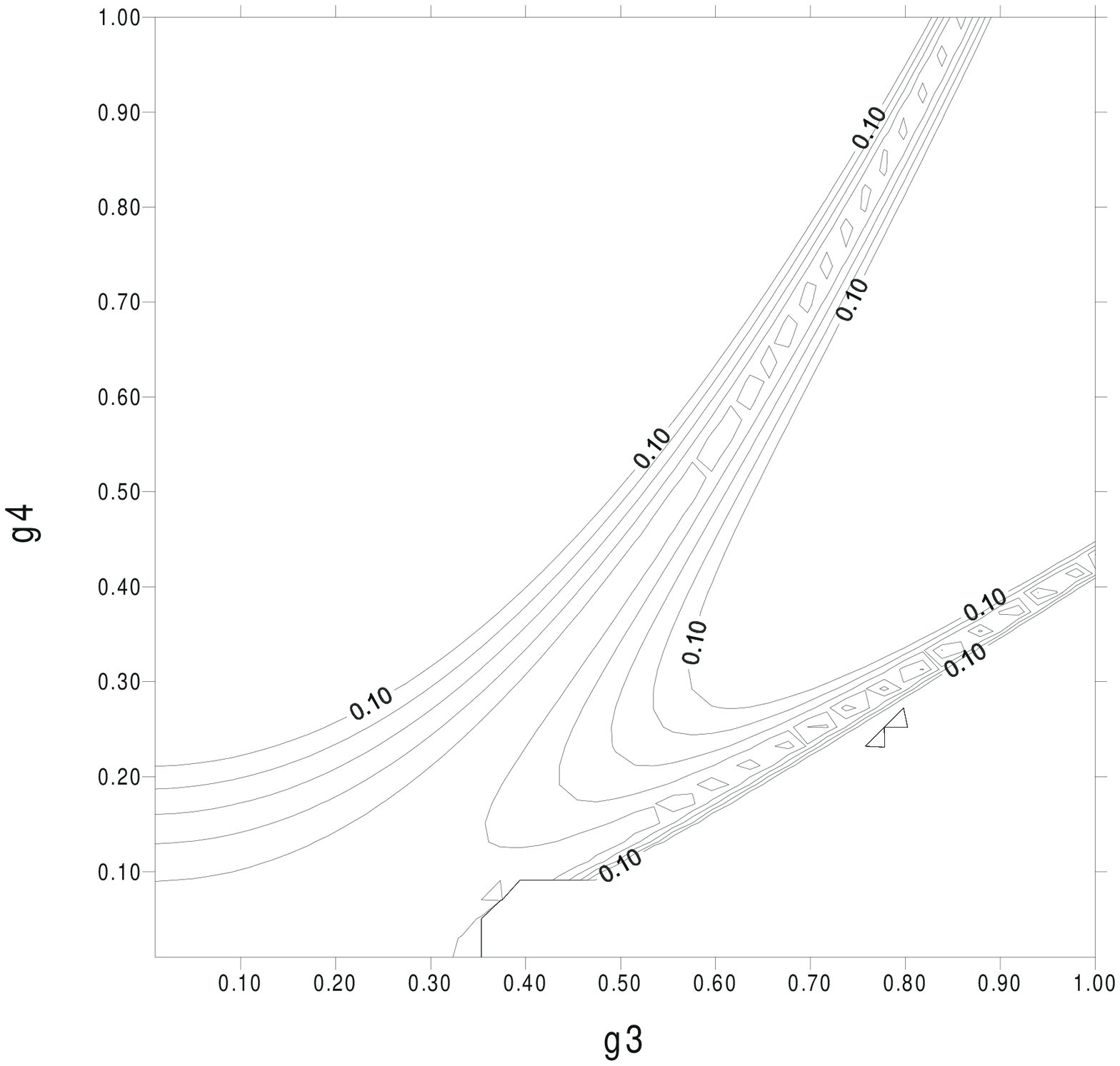}
    \includegraphics[height=7cm]{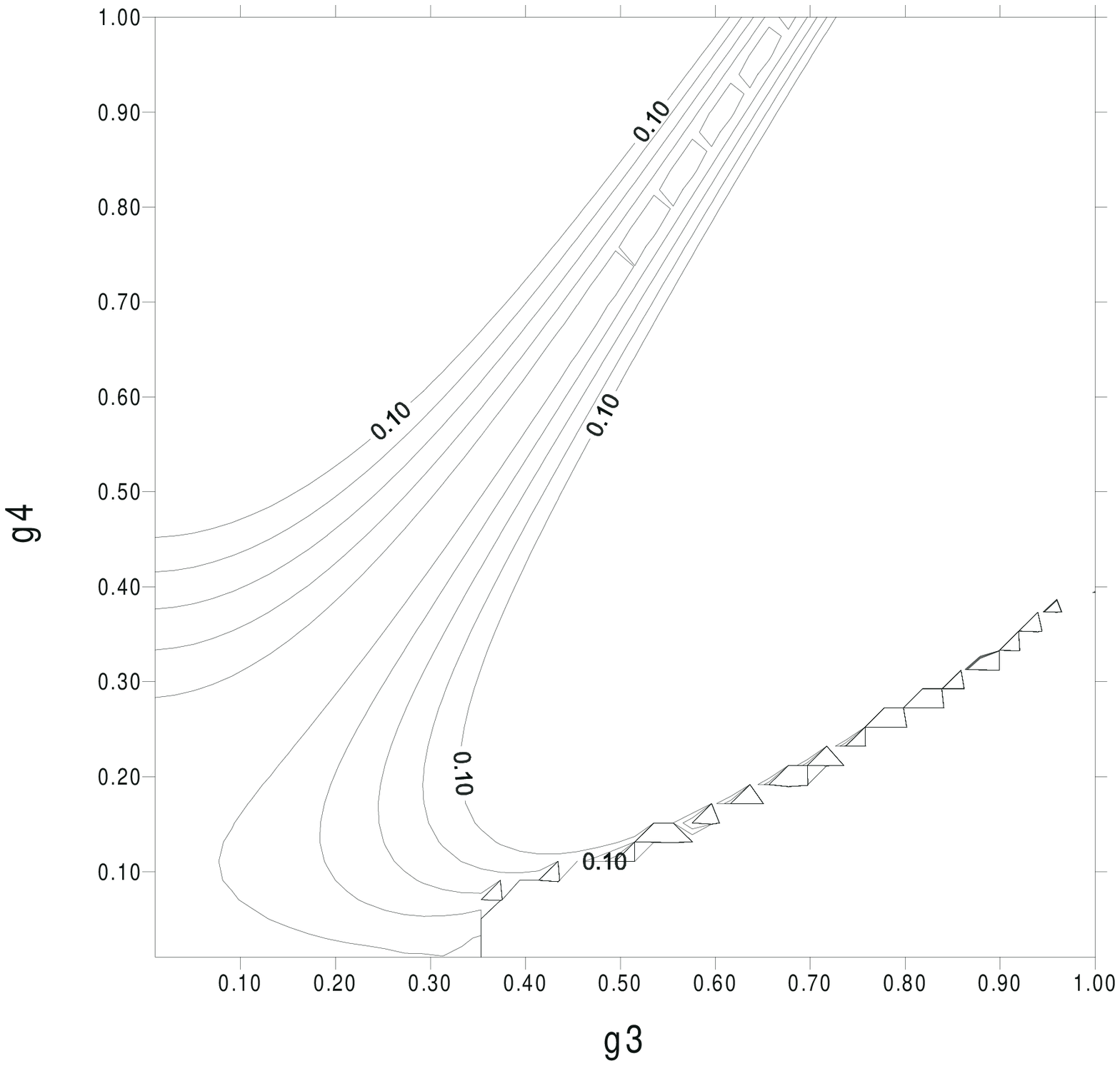}
    \caption{$|\delta G_1^\mathrm{1-loop}|<0.1$ (top left),
    $|\delta G_1^\mathrm{2-loop}|<0.1$ (top right),
    $|\delta G_1^\mathrm{improved}|<0.1$ (bottom)}
    \label{g1}
\end{figure}

All our calculations have been done for $g_4>\frac{3}{8}\,g_3^2$ where the
above action has a unique minimum at $\phi=0$. For the two-point function the
one-loop approximation is given in Figure~\ref{l1g2}.
In this plot we see the expected perturbative region in the vicinity
of the origin. What is not immediately obvious is the meaning of the
two regions sprouting off to large values of the coupling constants.
Still, it is rather easy to give a simple hand-waving argument: The
one-loop result for $G_n$ is an $(n-1)^\mathrm{st}$ order polynomial
in the couplings, while the exact result is a relatively slowly varying,
monotonous function (on the range of interest). As a consequence,
$\delta G_n^\mathrm{1-loop}$ can vanish on (at most) $n$ curves in
the $g_3,g_4$ plane. The sprouting regions in Figure~\ref{l1g2} flank
these two curves. A similar plot of $\delta G_1^\mathrm{1-loop}$ has
one sprouting region. What is important to note is that the sprouting
regions corresponding to different Green's function have nothing to do
with each other. In fact, they are just a
manifestation of the old saw that a stopped clock gives the correct
time of day twice during each day. The same plot for the two-loop and
improved Gaussian approximations is shown in Figure~\ref{l2g2+ig2}. 
As before, we again have central (perturbative) regions and a certain number
of sprouting regions. These plots clearly show that the improved Gaussian
approximation outperforms both the one-loop and two-loop results.
Figure~\ref{g1} gives the same comparison of one-loop, two-loop and
improved results for the one-point Green's function.

We have looked at all the Green's function from $G_1$ to $G_6$. In all
cases the improved Gaussian approximation gives the best result, however
the advantage becomes less marked when one looks at higher Green's function.

As a conclusion, we have shown in this paper that even the simplest
improved Gaussian approximation gives better agreement with exact results
than both one-loop and two-loop approximations.
At the same time, the computational cost of the
improved Gaussian approximation is negligibly greater than that of the
one-loop result, and significantly smaller than of the two-loop result.
In our previous paper we looked at the benefits of the improved Gaussian
approximation both from the analytic and numerical (Monte Carlo) sides.
We are currently working \cite{new} on extending those results to
interesting models in $d\ge 1$.

\end{document}